\documentclass[preprint,onecolumn]{rmaa}

\usepackage{supertabular}

\title{Quantitative Stellar Spectral\\
       Classification. III. Spectral Resolution}

\author{J. Garc\'{\i}a\altaffilmark{1,2},
        J. Stock\altaffilmark{3}\,$^\dag$,
        M. J. Stock\altaffilmark{1,3,4} and
        N. S\'anchez\altaffilmark{1,5}}

\altaffiltext{1}{Laboratorio de Investigaciones Astron\'omicas,
                 Departamento de F\'{\i}sica,
                 Universidad del Zulia,
                 Venezuela.}
\altaffiltext{2}{Postgrado de F\'{\i}sica Fundamental,
                 Centro de F\'{\i}sica,
                 Instituto Venezolano de Investigaciones Cient\'{\i}ficas
                 (IVIC), Venezuela.}
\altaffiltext{3}{Centro de Investigaciones de Astronom\'{\i}a (CIDA),
                 Venezuela.}
\altaffiltext{4}{Postgrado de F\'{\i}sica Fundamental,
                 Facultad de Ciencias,
                 Universidad de Los Andes,
                 Venezuela.}
\altaffiltext{5}{Instituto de Astrof\'{\i}sica de Andaluc\'{\i}a,
                 Consejo Superior de Investigaciones Cient\'{\i}ficas (CSIC), 
                 Espa\~na.\newline $^\dag$ Deceased April 19th 2004}

\fulladdresses{
\item Javier Garc\'{\i}a:
      Centro de F\'{\i}sica, Instituto Venezolano de Investigaciones
      Cient\'{\i}ficas (IVIC), Caracas 1020A, Venezuela
      (jgarcia@ivic.ve)
\item M. Jeanette Stock and
      N\'estor S\'anchez:
      Laboratorio de Investigaciones Astron\'omicas,
      Departamento de F\'{\i}sica, 
      Facultad Experimental de Ciencias,
      Universidad del Zulia,
      Apartado 15439, Maracaibo, Venezuela
      (mjstock@cida.ve,nestor@iaa.es)}

\shortauthor{Garcia et al.}
\shorttitle{Quantitative Stellar Spectral Classification. III.}

\resumen{
El m\'etodo desarrollado por Stock \& Stock (1999) para
derivar magnitudes absolutas y colores intr\'{\i}nsecos se
aplica a espectros simulados de baja resoluci\'on.
La simulaci\'on se realiza convolusionando espectros
verdaderos con una funci\'on Gaussiana, donde $\sigma$ (el ancho
a la altura media) est\'a relacionado con la
resoluci\'on final del espectro. La precisi\'on con
la cual se determinan los par\'ametros estelares indica
que este m\'etodo puede ser aplicado a
espectros t\'{\i}picos de prisma objetivo. Sin embargo,
los resultados muestran que las variaciones en la
resoluci\'on espectral no afectan los par\'ametros
derivados s\'olo para las estrellas m\'as tempranas,
mientras que para las estrellas de tipo m\'as tard\'{\i}as
se hace necesario mejorar el m\'etodo aplicado.
}

\abstract{
The method developed by Stock \& Stock (1999) to derive
absolute magnitudes and intrinsic colors is applied to
simulated low-resolution spectra. The simulation is made by
convolving real spectra with a Gaussian function, $\sigma$ (the
full width at half maximum), related to the final spectral
resolution. The accuracy with which the stellar parameters
are determined indicates that the method may be applied
to typical objective-prism spectra.
We show that changes in the spectral resolution
do not significantly affect the stellar parameters obtained
with this method for early-type stars, whereas for
later-type stars an improved approach is necessary.
}


\begin{document}
 
\maketitle                                                                      

\section{Introduction}

In a previous work, Stock \& Stock (1999, hereafter Paper I)
developed a quantitative method to obtain stellar physical
parameters such as absolute magnitude, intrinsic color, and
a metallicity index. These parameters were calculated by the use
of pseudo-equivalent widths of
absorption features in stellar spectra by means of polynomials.
For this purpose, the spectral library of Jones (1999, available
at ftp://ftp.noao.edu/catalogs/coudelib) was used
which contains about 700 stellar spectra in the bands 3820-4500
\AA\ and 4780-5450 \AA\, with a resolution FWHM of 1.8 \AA. The
calibration (i.e. the determination of the coefficients of the
polynomials) was made using 487 stellar spectra
of A-K types. In such spectra, the
definition of the true continuum is imprecise. For this reason,
in Paper I, the authors worked with pseudo-equivalent widths 
for the identified
features. This was done by selecting one inner and two outer
regions for each absorption line. The two outer regions (one
on each side of the line) were used to determine the
pseudo-continuum, while the integration of the inner
region allows us to know the pseudo-equivalent width. This
procedure is similar to that used by Worthey et al. (1994).

One of the main goals of the quantitative classification 
method is its application to a large number of
stellar spectra, obtained by means of objective-prism
observations, which are a kind of low resolution spectra.
Therefore, it is essential to simulate
low-resolution spectra. This may be done by applying a mathematical 
treatment to smooth the spectra of Jones library. Once this is done,
it is necessary to re-calibrate the method using those smoothed
spectra in order to compare the results with those obtained
in Paper I. It is important to know if the method is as
accurate as in Paper I when it is used with low resolution
spectra.

On the other hand, the equivalent width of a line does
not depend, in principle, on the spectral resolution.
Considering the way that the pseudo-equivalent widths were
defined in Paper I, these could be affected
by variations in the spectrum profile. It is also important
to analyze the sensitivity of the method to variations
in the spectral resolution (due, for instance, to seeing
variations during the observations). To do this, first the
polynomials are generated applying the method to the smoothed
spectra. Then these are applied to spectra of two other resolutions, 
and the respective results are compared.

A short explanation of how the low-resolution spectra are simulated
and how the method is applied to these is given in Sections~2 and 3.
Section~4 is dedicated to a
discussion of the results, and the main conclusions are
summarized in Section~5.

\section{Low-resolution spectra simulation}

The smoothing method consists of taking each pixel
and then distributing its intensity among the neighboring
pixels using a mathematical function. The spectra of the
Jones library were convolved with a Gaussian function,
$\sigma$ (the full width at half maximum) being the
smoothing parameter. The number of neighboring pixels
used in this process increases as $\sigma$ increases
with the consequent decrease of the spectral details
(lower resolution). All the spectra of Jones library
were smoothed using 14 different values of $\sigma$
from 20 to 700.

As was pointed out in Paper I, objective-prism
spectra are very useful in galactic structure
studies which is the main motivation for the
development of this quantitative classification
method. Even though the objective-prism spectral
dispersion is not a linear function of $\lambda$ (wavelength),
we simulate the low-resolution spectra using a linear
relationship. This might be done since the main goal of 
this work is to
analyze the sensitivity of the method to changes
in the spectral resolution. Both the pseudo-equivalent
widths and the line depths depend on the spectral
resolution. Thus, we defined an index as the ratio
of these two quantities in order to compare smoothed spectra with real
objective-prism spectra. In this way, we could identify
which one of the indices better matches the observations.
For early-type stars, the hydrogen
H$\gamma$ and H$\delta$ lines were selected whereas for
late-type stars the G-band was used.

\begin{figure}[t]
\begin{center}
\includegraphics[width=\textwidth]{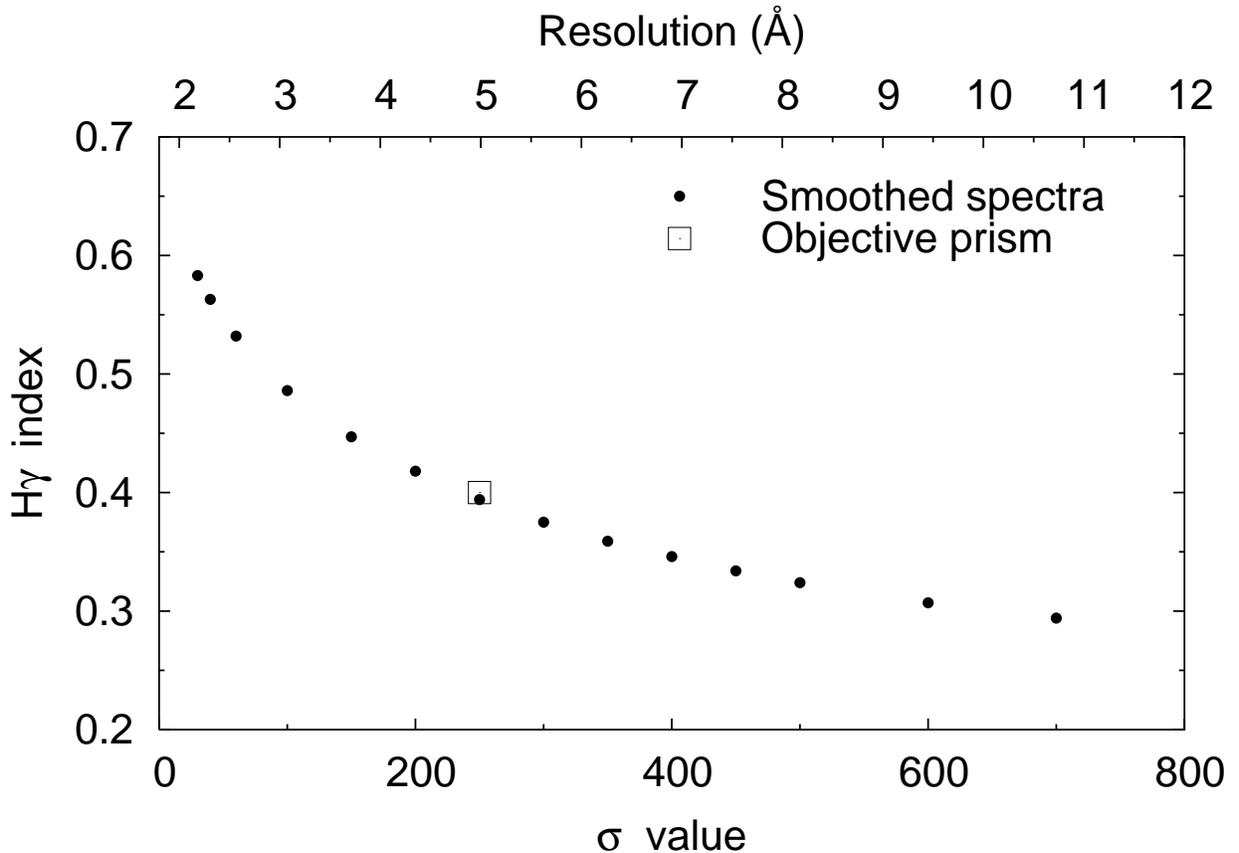}
\caption{Variation of the (pseudo-equivalent
         width/line depth) ratio with the smoothing
         parameter $\sigma$ (bottom axis) or the
         smoothed resolution (top axis) for the hydrogen
         line H-$\gamma$. The open square indicates the
         value for a typical objective-prism spectrum.}
\label{hgamma}
\end{center}
\end{figure}

For the resolutions corresponding to all $\sigma$ values,
these indices (H$\gamma$, H$\delta$, and G-band) were
calculated on typical A0V and G0 stars from the
Jones library.
Additionally, the same indices were calculated for
typical objective-prism spectra of A0V and G0
spectral types obtained by Stock (1997) with the
60 cm Curtis Schmidt telescope at CTIO (Cerro Tololo
Interamerican Observatory), using a 4-degree prism
and a B filter. As an example, Figure~\ref{hgamma}
shows the H$\gamma$ index for an early-type
spectrum taken from the Jones library as a
function of both $\sigma$ (bottom X-axis) and
the final smoothed resolution in angstrom
(top X-axis). The open square indicates the
value of this index for a typical objective-prism
spectrum. In this way, we were able to determine that
$\sigma \simeq 250$ reproduces the resolution ($\simeq
5$ \AA) that best fits objective-prism observations.

\begin{figure}[t]
\begin{center}
\includegraphics[width=\textwidth]{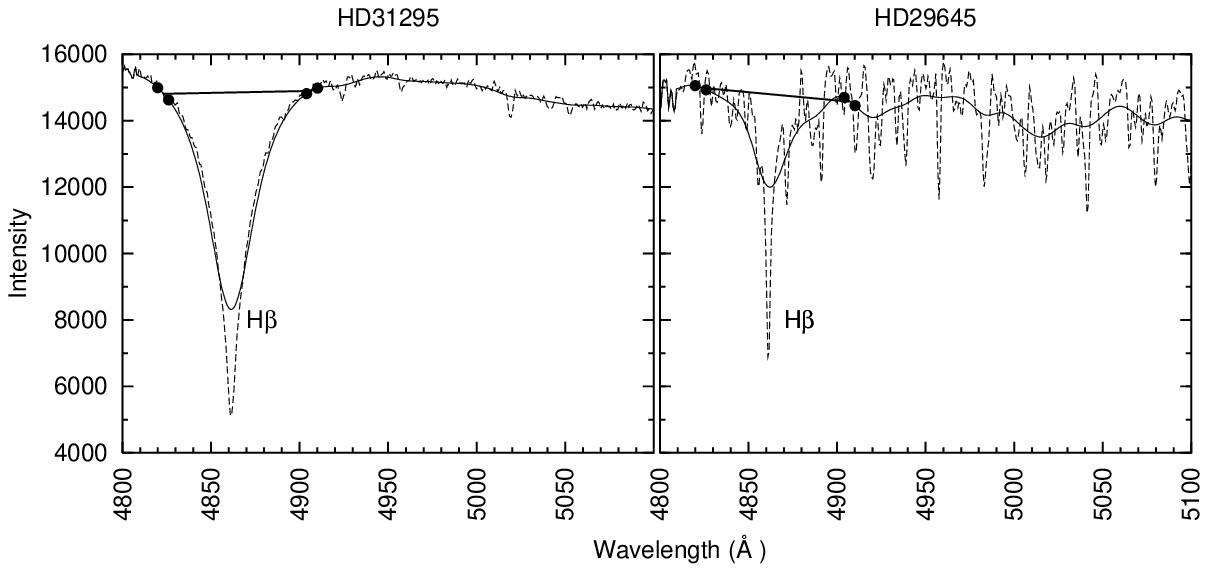}
\caption{The spectra of HD31295 (left-side panel) and HD29645
         (right-side panel) from the Jones library (dashed
         lines) and the corresponding simulations for $\sigma =
         250$ (continuous lines). The black points indicate the
         ranges defining an example line (H$\beta$) and the
         continuum regions (see text).}
\label{jones}
\end{center}
\end{figure}

In addition to this value, that we will name the
``standard resolution", we chose two other values of
$\sigma$ (150 and 350) to simulate higher and lower
spectral resolutions, respectively. The reason for
this is to estimate the sensitivity of the method to
variations in the resolution. This was carried out
by applying the polynomials obtained with the standard
resolution to the other two resolutions and comparing
the results. As an example, Figure~\ref{jones}
shows the spectra of both an early-type star (HD31295,
type A0V) and a late-type star (HD29645, type G0V), taken from
the Jones library. The corresponding simulations for
$\sigma = 250$ (standard resolution) are also showed where
the changes in the shape of the line profiles can be seen.
Solid circles in this Figure indicate windows defining
a given line (in this case H$\beta$) in the smoothed spectra,
and the both-sides continuum regions on which we average the
fluxes to define the pseudo-continuum. The solid line shows
the resulting pseudo-continuum used to calculate the
pseudo-equivalent width for this line.

\section{Application of the method}

The determination of the equivalent widths and the
calculation of absolute magnitudes and color indices
$(B-V)_0$ were performed by the use of the same codes developed
in Paper I. However, the points that define the
pseudo-continuum for each identified feature were
properly re-defined for the spectra smoothed with
the standard resolution. This was done because the features of the
original spectra are very different from those of the
smoothed ones as can be seen in Figure~\ref{jones}. The
new points and the identified lines are contained in
Table~\ref{lines}.
\begin{table}
\caption{List of the lines selected in Paper I\\
         and the re-defined points}
\label{lines}
\begin{center}
\begin{tabular}{ccccccccc}\hline\hline
Line  & \multicolumn{2}{c}{Cont1(\AA)} & 
\multicolumn{2}{c}{Line(\AA)} & 
\multicolumn{2}{c}{Cont2(\AA)} & 
$\lambda$ (\AA)  &  Identification \\
\hline
   1 & 3859.9 & 3863.0 & 3863.0 & 3910.3 &
       3910.3 & 3915.3 & 3887.9 & H$\psi$+HeI  \\
   2 & 3910.3 & 3915.3 & 3915.3 & 3950.2 &
       3950.2 & 3953.3 & 3934.6 & CaII K \\
   3 & 3915.9 & 3950.2 & 3950.2 & 4003.1 & 
       4003.1 & 4009.3 & 3968.9 & CaII H + H$\epsilon$ \\
   4 & 4009.3 & 4018.7 & 4018.7 & 4046.7 & 
       4046.7 & 4052.9 & 4031.8 & FeI \\
   5 & 4043.6 & 4056.0 & 4056.0 & 4143.3 & 
       4143.3 & 4158.8 & 4102.1 & H$\delta$ \\
   6 & 4183.8 & 4205.6 & 4205.6 & 4246.1 & 
       4246.1 & 4264.7 & 4227.4 & CaI \\
   7 & 4274.1 & 4280.3 & 4280.3 & 4323.9 & 
       4323.9 & 4333.3 & 4304.0 & G-band \\
   8 & 4317.7 & 4321.4 & 4322.7 & 4328.3 & 
       4328.9 & 4335.2 & 4325.8 & FeI, CH \\
   9 & 4280.3 & 4317.7 & 4317.7 & 4365.1 & 
       4365.1 & 4420.5 & 4340.8 & H$\gamma$ \\
  10 & 4358.2 & 4364.4 & 4364.4 & 4411.2 & 
       4411.2 & 4423.6 & 4386.2 & FeI \\
  11 & 4420.5 & 4445.4 & 4445.4 & 4473.5 & 
       4473.5 & 4476.6 & 4457.9 & Blend (CaI, MnI, FeI) \\
  12 & 4819.9 & 4826.1 & 4826.1 & 4904.0 & 
       4904.0 & 4910.2 & 4862.9 & H$\beta$ \\
  13 & 4904.0 & 4910.2 & 4910.2 & 4947.6 & 
       4947.6 & 4953.8 & 4918.9 & FeI \\
  14 & 4975.6 & 4994.3 & 4994.3 & 5028.6 & 
       5028.6 & 5059.7 & 5013.6 & -- \\
  15 & 5028.6 & 5059.7 & 5059.7 & 5090.9 & 
       5090.9 & 5122.1 & 5079.1 & Blend (FeI, NiI) \\
  16 & 5090.9 & 5159.4 & 5159.4 & 5196.8 & 
       5196.8 & 5215.5 & 5175.0 & MgI \\
  17 & 5234.2 & 5243.5 & 5243.5 & 5287.2 & 
       5287.2 & 5312.1 & 5267.2 & Blend (CaI, FeI) \\
  18 & 5287.2 & 5312.1 & 5312.1 & 5358.8 & 
       5358.8 & 5383.7 & 5328.9 & FeI \\
  19 & 5358.8 & 5383.7 & 5383.7 & 5421.1 & 
       5421.1 & 5433.6 & 5403.7 & FeI \\
\hline
\end{tabular}
\end{center}
\end{table}
For all these lines, the equivalent widths were determined,
and polynomials up to the second degree were defined with
three lines as independent variables.
The stars were divided into groups as in Paper I, i.e.
1=very early stars, 2=early stars, 3=late stars and
4=very late stars. The separation into groups is made
in basis of the number of lines falling within certain
predefined ranges of equivalent widths,
and it does not correspond to any
specific spectral types or colors (see Figure 6 in
Paper I), therefore the adjectives ``early" or ``late"
must be taken carefully. The range of spectral types
is approximately B9-A2 for stars belonging to the group
1, F0-G5 for group 2, G0-K0 for group 3 and G5-M0 for
group 4.
Nevertheless, group 1 was excluded from the analysis
because the number of available spectra was small
since we only took into account stars
for which the Hipparcos catalogue gives parallaxes
with an error less than 20\%.

Afterwards, the magnitudes and $(B-V)_0$ colors for
spectra with $\sigma = 150$ and 350 were
calculated by using the polynomials obtained
for each group with the standard resolution spectra.
It is important to
emphasize here that the polynomials and the
coefficients used to calculate absolute magnitudes
and colors are, for all the three resolutions,
those derived from the standard resolution. This
is necessary in order to analyze the sensitivity
of the method to the spectral resolution.

\section{Analysis}

For the standard resolution ($\sigma = 250$), the
best solutions obtained with combinations of three
lines are shown in Table~\ref{best} (N$_u$ indicates
the number of stars used),
\renewcommand{\baselinestretch}{1}
\begin{table}
\caption{Best combinations for the determination of the
         stellar parameters with the standard resolution}
\label{best}
\begin{center}
\begin{tabular}{c}
Absolute Magnitude
\end{tabular} \\
\begin{tabular}{ccccccccccccccc}\hline\hline
& \multicolumn{3}{c}{Group 2} & & & \multicolumn{3}{c}{Group 3} 
& & & \multicolumn{3}{c}{Group 4} &  \\
% \hline
L1 & L2 & L3 & $rms$ & N$_u$ &  L1 & L2 & L3 & $rms$ 
& N$_u$ & L1 & L2 & L3 & $rms$ & N$_u$  \\
\hline
   1 &    3 &   16 &  0.273 &   153 &    1 &    8 &
   12 &  0.256 &   143 &    1 &    2 &   18 &  0.263 &   115 \\
   1 &    4 &   16 &  0.274 &   144 &    1 &    9 &
   16 &  0.222 &   148 &    1 &    3 &   14 &  0.248 &   116 \\
   3 &   10 &   16 &  0.288 &   153 &    1 &   12 &
   16 &  0.286 &   155 &    1 &    9 &   14 &  0.222 &   114 \\
   3 &   13 &   16 &  0.274 &   152 &    2 &    6 &
    8 &  0.223 &   143 &    2 &    3 &   16 &  0.254 &   123 \\
   9 &   16 &   19 &  0.273 &   154 &    2 &    6 &
   16 &  0.237 &   159 &    2 &    5 &   16 &  0.245 &   119 \\
  10 &   14 &   16 &  0.253 &   155 &    2 &    9 &
   16 &  0.277 &   162 &    2 &    6 &   10 &  0.264 &   121 \\
  12 &   14 &   16 &  0.273 &   156 &    5 &    7 &
   16 &  0.269 &   163 &    2 &    7 &   16 &  0.243 &   121 \\
  12 &   16 &   19 &  0.292 &   153 &    5 &    9 &
   16 &  0.264 &   157 &    2 &   12 &   16 &  0.246 &   120 \\
  13 &   14 &   16 &  0.258 &   155 &    5 &   13 &
   16 &  0.256 &   158 &    2 &   13 &   16 &  0.254 &   117 \\
  16 &   18 &   19 &  0.286 &   155 &    7 &    9 &
   16 &  0.282 &   165 &    3 &    7 &   16 &  0.260 &   124 \\
\hline
\end{tabular} \\
\vspace{1cm}
\begin{tabular}{c}
Intrinsic Color $(B-V)_0$
\end{tabular} \\
\begin{tabular}{ccccccccccccccc}\hline
 & \multicolumn{3}{c}{Group 2} & & & \multicolumn{3}{c}{Group 3} 
& & & \multicolumn{3}{c}{Group 4} &  \\
% \hline
L1 & L2 & L3 & $rms$ & N$_u$ &  L1 & L2 & L3 & $rms$ & N$_u$ 
& L1 & L2 & L3 & $rms$ & N$_u$  \\
\hline
   1 &    3 &   14 &  0.019 &   143 &    3 &    5 &
    7 &  0.018 &   157 &    1 &    2 &    3 &  0.023 &   118 \\
   1 &    4 &    7 &  0.018 &   143 &    3 &    5 &
   14 &  0.019 &   158 &    2 &    3 &   10 &  0.023 &   118 \\
   1 &    5 &    6 &  0.018 &   143 &    4 &    5 &
    6 &  0.022 &   161 &    2 &   10 &   19 &  0.024 &   114 \\
   1 &    6 &   10 &  0.015 &   136 &    5 &    7 &
   13 &  0.021 &   157 &    3 &    4 &   13 &  0.024 &   124 \\
   2 &    6 &    9 &  0.018 &   142 &    5 &    7 &
   17 &  0.019 &   153 &    3 &    5 &    6 &  0.018 &   120 \\
   2 &    7 &   14 &  0.019 &   144 &    5 &    7 &
   19 &  0.019 &   152 &    3 &    5 &    7 &  0.022 &   122 \\
   4 &    8 &    9 &  0.018 &   144 &    5 &    8 &
   12 &  0.019 &   152 &    3 &    5 &   12 &  0.021 &   121 \\
   6 &    9 &   17 &  0.018 &   145 &    5 &    8 &
   14 &  0.021 &   155 &    3 &    5 &   15 &  0.020 &   120 \\
  10 &   14 &   16 &  0.018 &   144 &    5 &   10 &
   16 &  0.019 &   153 &    3 &    6 &   10 &  0.022 &   121 \\
  13 &   14 &   17 &  0.019 &   145 &    5 &   14 &
   19 &  0.022 &   159 &    3 &   10 &   17 &  0.022 &   119 \\
\hline
\end{tabular}
\end{center}
\end{table}
\renewcommand{\baselinestretch}{1.5}
both for the absolute
magnitude and for the $(B-V)_0$ color. We see that
absolute magnitudes can be recovered with an average
error of about 0.26 magnitudes, and $(B-V)_0$ colors
with 0.020 magnitudes for all the groups. Thus, the
results obtained with the standard resolution are in 
agreement with the results obtained in Paper I,
the basic difference being that the smoothed spectra
simulate properly objective-prism spectra within the
assumptions explained in Section~2.

\begin{figure}[t]
\begin{center}
\includegraphics[width=\textwidth]{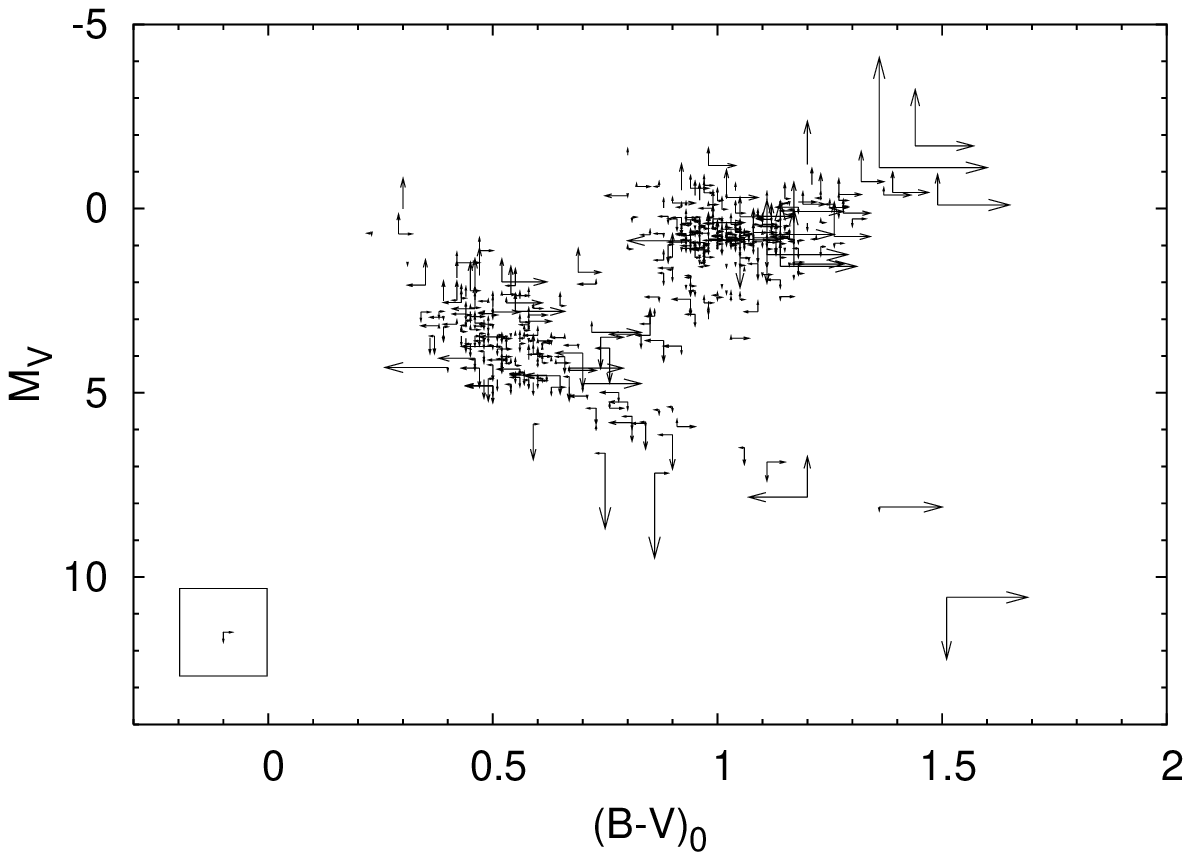}
\caption{Color-magnitude diagram of the stars used 
         from the Jones library and the Hipparcos catalogue.
         The length of each arrow is the difference between
         the stellar parameters calculated in Paper I
         and the parameters calculated with
         $\sigma = 250$.}
\label{hr}
\end{center}
\end{figure}

The intrinsic color-magnitude diagram of the stars
used in this work is shown in Figure~\ref{hr}.
The lengths of the arrows are the differences between
the values of the physical parameters determined in
Paper I, using the full-resolution spectra, and those
determined in the present work for the same points,
but using the spectra smoothed with the standard
resolution. As reference, the arrows in the box
indicate the mean rms of the method for
$M_{V}$ and $(B-V)_0$. Generally, the lengths
of the arrows are small, except for a few late-type
stars where the solution is not very robust because
of the lack of data in that region.

\begin{figure}[t]
\begin{center}
\includegraphics[width=\textwidth]{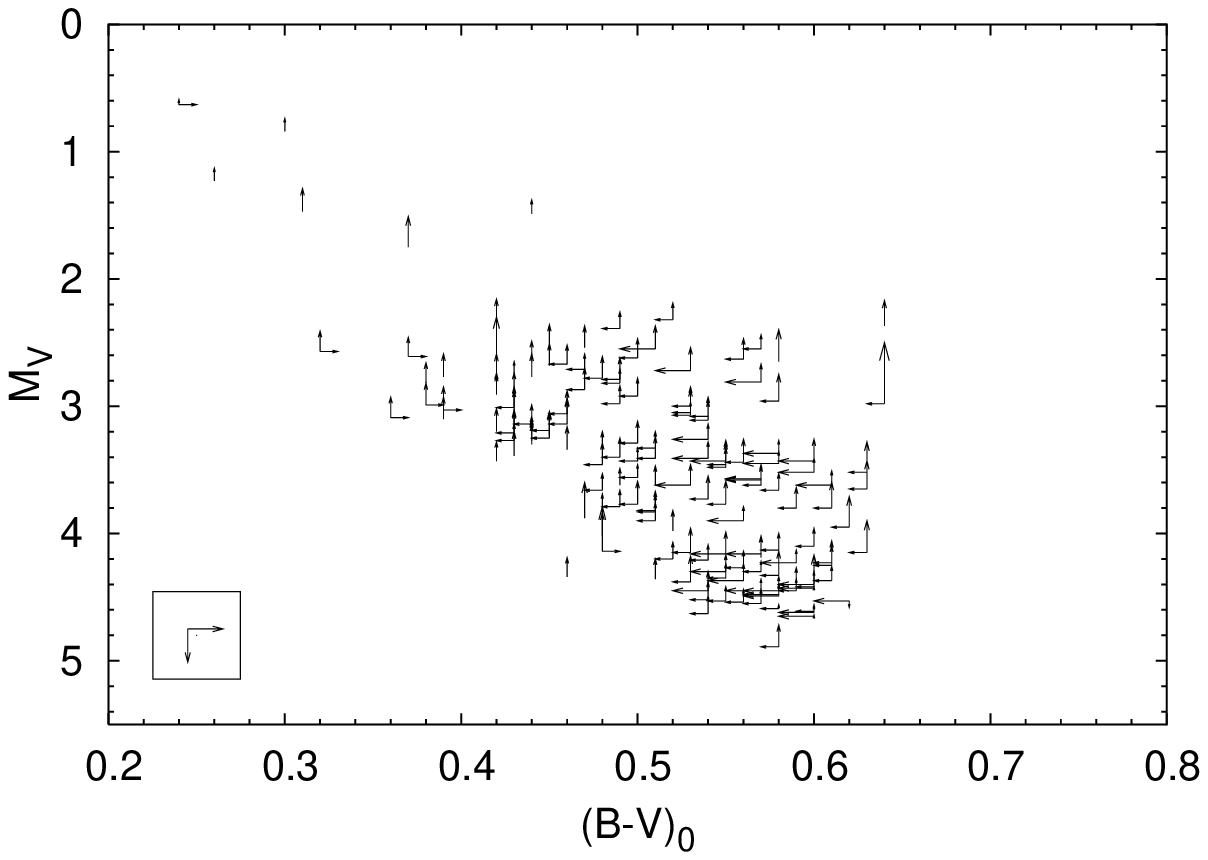}
\caption{Color-magnitude diagram obtained from the
         calibration with $\sigma = 250$ (standard
         resolution), for the stars of group 2.
         The length of each arrow is the difference
         between the stellar parameters calculated
         with $\sigma = 250$ and the parameters
         calculated with $\sigma = 350$.}
\label{g2-350}
\end{center}
\end{figure}

For each group of stars, the effect of spectral resolution
variations was analyzed by comparing the obtained stellar
parameters for the spectra smoothed with different values of
$\sigma$. As was pointed out before, the same polynomial
should be used for all the smoothed spectra and for this reason
we chose the polynomial obtained for the standard resolution
with the best combination of lines (see Table~\ref{best}). 
Figure~\ref{g2-350}
is a color-magnitude diagram for the stars belonging
to group 2. In this case, the origin of each arrow
is given by the magnitudes and intrinsic colors obtained
with the standard resolution ($\sigma = 250$), whereas
the lengths of the arrows are given by the differences
between the values of the physical parameters calculated
for the spectra smoothed with $\sigma = 350$ and $\sigma 
= 250$. A trend to decrease $(B-V)_0$
when the spectral resolution is decreased can be noted (i.e. when 
$\sigma$ is increased), although the changes in the 
intrinsic color are always smaller than the mean error of
the adjustment of the applied polynomial. In fact, the
averaged value of the difference of colors (between both
resolutions) is -0.008 magnitudes with a rms of 0.008 
magnitudes, whereas the mean rms resulting from the
application of the method is, as mentioned 
above, 0.020 magnitudes. The observed tendency in the
absolute magnitude is that $M_{V}$ gets brighter as the resolution
decreases. The mean value of the differences in magnitude
is -0.15 magnitudes (with a rms of 0.06 magnitudes) which
is smaller than the mean rms of the solution (0.26
magnitudes). Thus, even when a systematic behavior is
observed both in magnitude and color, this effect is
not significant.

\begin{figure}[t]
\begin{center}
\includegraphics[width=\textwidth]{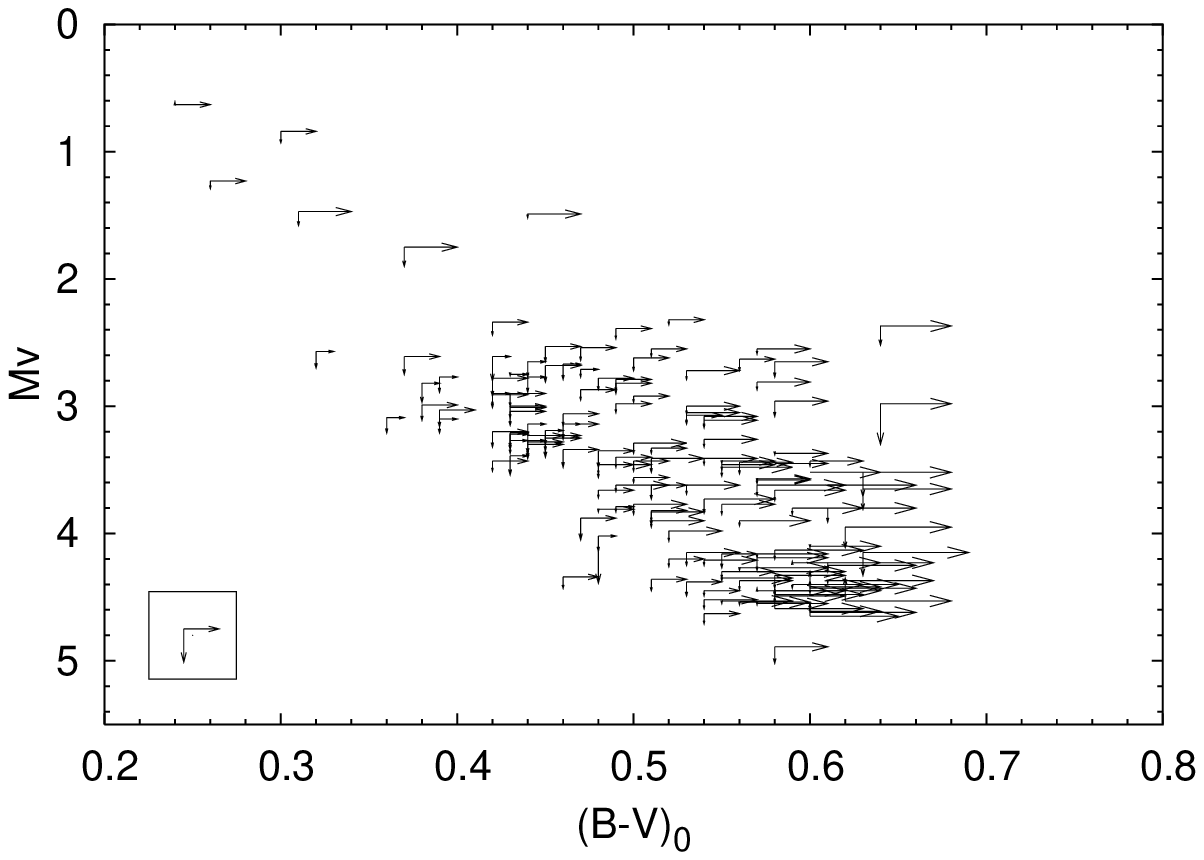}
\caption{Color-magnitude diagram obtained from the
         calibration with $\sigma = 250$ (standard
         resolution), for the stars of group 2.
         The length of each arrow is the difference
         between the stellar parameters calculated
         with $\sigma = 250$ and the parameters
         calculated with $\sigma = 150$.}
\label{g2-150}
\end{center}
\end{figure}

Figure~\ref{g2-150}
shows a color-magnitude diagram
for the stars of the same group (2), but now
the stellar parameters determined for the spectra smoothed
with $\sigma = 250$ (standard resolution) and $\sigma =
150$ (higher resolution) are compared. As the resolution
increases, $(B-V)_0$ increases, but an additional effect is
also observed: generally, the increase in the color
is higher for the later-type stars of this group. However,
the mean value of the differences in color between both
resolutions is 0.028 magnitudes (with a rms of 0.014
magnitudes), which is slightly higher than the rms of
the method. The $M_{V}$ magnitude gets fainter as resolution
increases, and in this case the differences in $M_{V}$ between
both resolutions is about 0.07 magnitudes with a rms
of 0.06 magnitudes (below the mean error of the method).

Therefore, from Figures~\ref{g2-350} and \ref{g2-150}
we conclude that for this group of stars, both the $(B-V)_0$
colors and the $M_{V}$ magnitudes tend to decrease when the
resolution of the spectra decreases. This systematic
behavior could be taken into account to correct it.
Nevertheless, we have seen that these changes are of the order
or less than the error with which the stellar parameters
can be recovered by the use of this method. Thus, the effects
of variations in the spectral resolution can be neglected,
at least when the method is applied to stars belonging to
the second group.

\begin{figure}[t]
\begin{center}
\includegraphics[width=\textwidth]{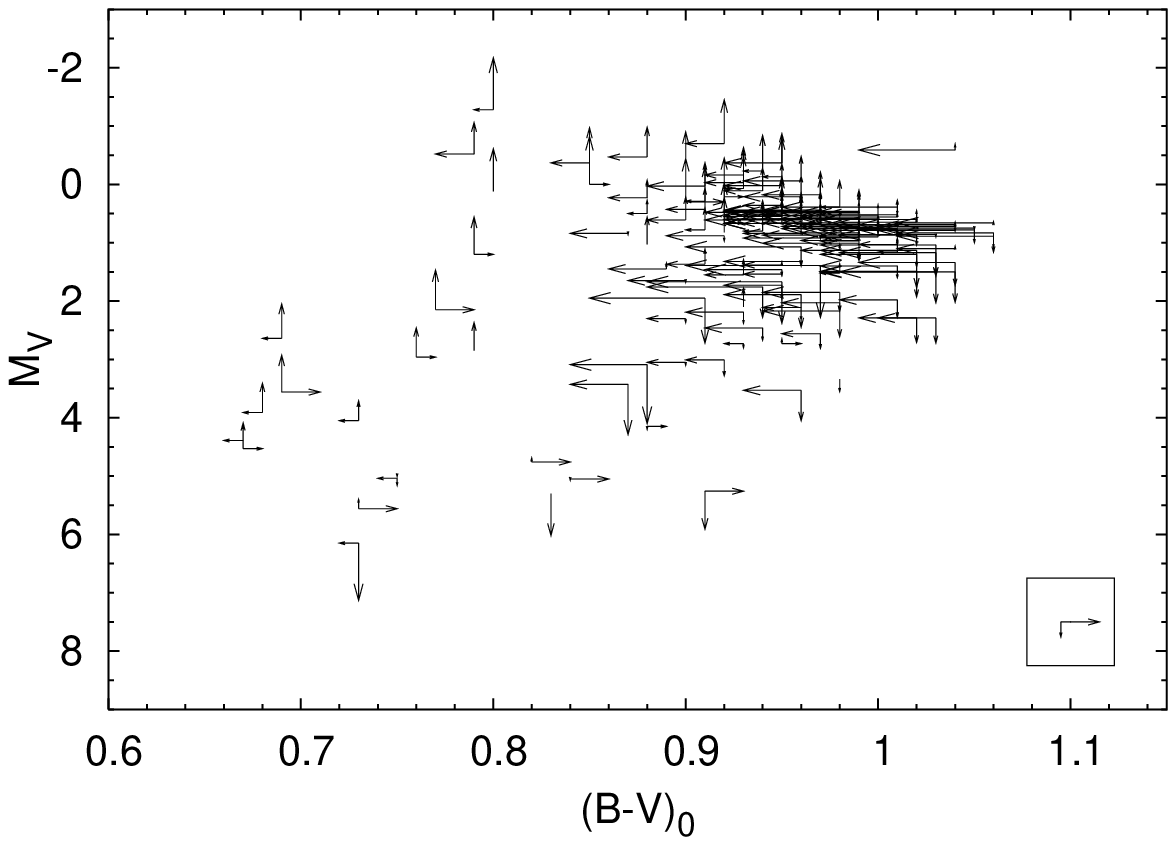}
\caption{Color-magnitude diagram obtained from the
         calibration with $\sigma = 250$ (standard
         resolution), for the stars of group 3.
         The length of each arrow is the difference
         between the stellar parameters calculated
         with $\sigma = 250$ and the parameters
         calculated with $\sigma = 350$.}
\label{g3-350}
\end{center}
\end{figure}

\begin{figure}[t]
\begin{center}
\includegraphics[width=\textwidth]{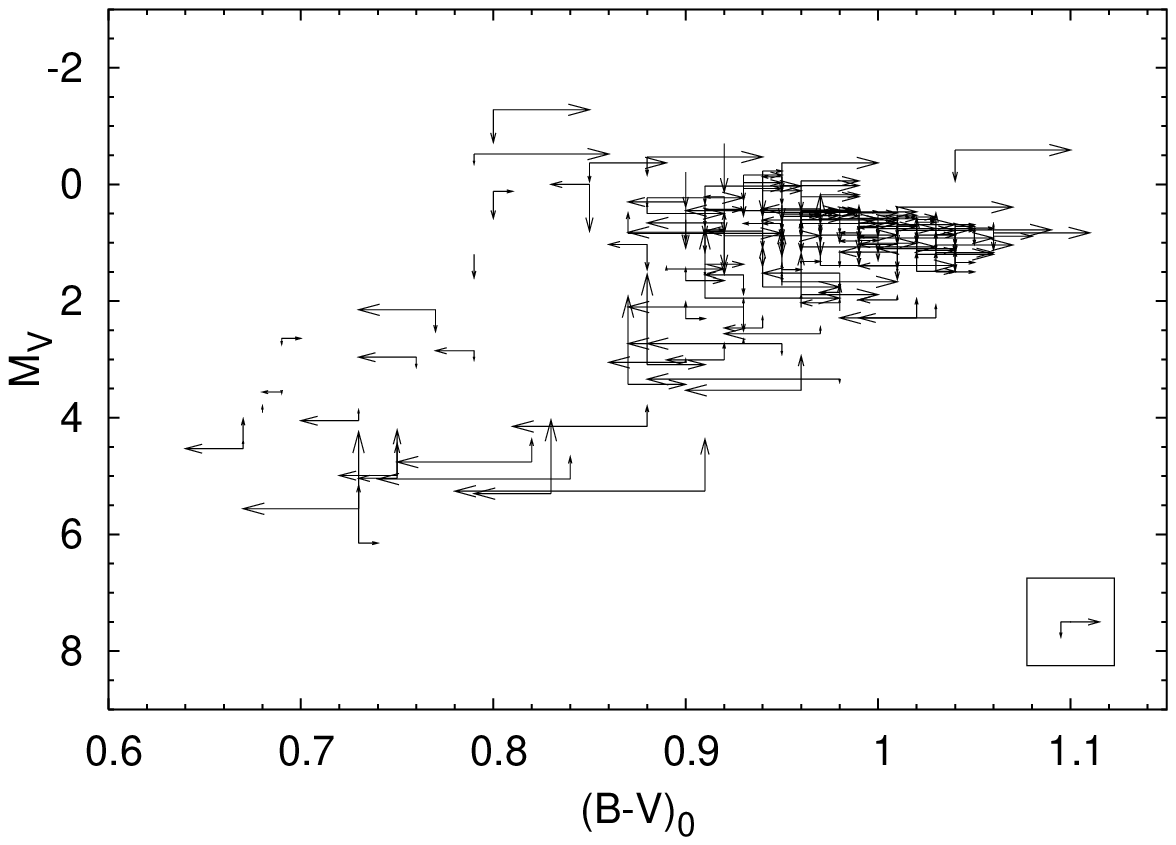}
\caption{Color-magnitude diagram obtained from the
         calibration with $\sigma = 250$ (standard
         resolution), for the stars of group 3.
         The length of each arrow is the difference
         between the stellar parameters calculated
         with $\sigma = 250$ and the parameters
         calculated with $\sigma = 150$.}
\label{g3-150}
\end{center}
\end{figure}

Regarding stars belonging to the group 3, the color-magnitude
diagram comparing the results obtained with $\sigma = 250$
with those obtained using $\sigma = 350$ is shown in
Figure~\ref{g3-350}. The corresponding comparison with
$\sigma = 150$ is shown in Figure~\ref{g3-150}.
The effect of variations in the resolution is more
perceivable in the stars of this group because of the
nature of this kind of spectra.
It can be noted that
there is no systematic effect when the resolution is changed,
i.e. some stars increase and other decrease $(B-V)_0$
(similarly for $M_{V}$). The averaged value of the difference
between the color determined with a given $\sigma$ value
and the color determined for $\sigma = 250$ are -0.026 mag
(rms = 0.021 mag) for $\sigma = 350$, and 0.005 mag (rms =
0.035 mag) for $\sigma = 150$. For the absolute
magnitudes, the results show
that the averages of the differences were -0.09 mag
(rms = 0.38 mag) and 0.09 mag (rms = 0.47) for $\sigma
= 350$ and $150$, respectively. Thus, for this group of stars,
we see that variations in the spectral resolution produce
changes in color and magnitude which are of the order or
even higher than the mean error of the adjustment
of the polynomial.

The differences between the colors and magnitudes
determined by using the spectra of various
resolutions become too high for very late stars
(group 4), rising up to 3 magnitudes for the absolute
magnitude and 1 magnitude for the $(B-V)_0$ color.
Obviously, this is due to the fact that these
stars have spectra with a great variety of
features and details, and the pseudo-continuum
of every absorption line cannot be
defined without ambiguity. In this case, the
method described in Paper I is not suitable to
obtain the stellar parameters with the desired 
accuracy, and another approach is
required.

\section{Conclusions}

There are two main conclusions drawn from the present
study. First, the application of the
quantitative classification
method (Paper I) to spectra smoothed with the standard
resolution ($\sigma = 250$), which simulates typical
objective-prism spectra, yielded results in good
agreement with the previous ones: it was possible
to derive absolute magnitudes with an average error
of 0.26 magnitudes and $(B-V)_0$ colors with an
error of 0.020 magnitudes. This method can then
be applied to objective-prism observations
with high accuracy.

Second, the results indicate that changes in the
resolution of the spectra will not affect
the determination of the measurable parameters
for the stars belonging to group 2 (earlier-type
stars). Concerning to very early stars (spectral types
O-B), the continuum can be easily determined and the
variations in the spectral resolution are not a
problem, neither for the calibration nor for the
derivation of the fundamental parameters (see Stock
et al. 2002). But, for later-type stars (groups
3 and 4), it is necessary to apply a different approach because
the shape of these spectra is very sensitive
to resolution variations. We propose that this group
should be divided into subgroups in order to
reduce the differences between the parameters
derived with several spectral resolutions.

\acknowledgments

The authors thank the referee for many helpful comments. 
It is a pleasure to thank Mr. Hender L\'{o}pez for his
collaboration in the revision of this paper.
This research has been partially supported by
CONDES of the Universidad del Zulia and by
FONACIT of Venezuela. 
N.S. acknowledges the financial support from the
Secretaria de Estado de Universidades e Investigacion
(Spain).

\end{document}